\documentstyle[preprint,aps]{revtex} 
\begin{document} 
\draft

\title{Radiation reaction in quantum mechanics}


\author{Atsushi Higuchi\footnote{e-mail address: ah28@york.ac.uk}}
\address{Department of Mathematics, University of York,\\ 
         Heslington, York YO10 5DD, United Kingdom}
\date{\today}
\maketitle 

\tightenlines
\begin{abstract}
The Lorentz-Dirac radiation 
reaction formula predicts that the position shift of a charged particle
due to the radiation 
reaction is of first order in 
acceleration if it undergoes a small acceleration.
A semi-classical calculation
shows that this is impossible at least if the acceleration is
due to a time-independent potential.  Thus, the
Lorentz-Dirac formula gives an incorrect classical limit in this
situation. The correct classical limit of the position shift at the lowest 
order in acceleration is obtained by assuming that the
energy loss at each time is given by the Larmor formula. 
\end{abstract} 

\narrowtext 


A classical charge $e$ with four-velocity $u^\mu$
emits electromagnetic
radiation with energy-momentum $P^{\mu}$ according to the Larmor formula:
\begin{equation}
\frac{dP^{\mu}}{d\tau} = \frac{2\alpha}{3} a^2\,u^{\mu}, \label{Larmor}
\end{equation}
where $\tau$ is the proper time along the world line of the charge $e$ and
where $a = \sqrt{-\dot{u}^{\mu}\dot{u}_\mu}$ (with 
$\dot{u}^\mu = d u^\mu/d\tau$) 
is the acceleration of the charge. (See, e.g., Ref.\ \cite{Jackson}.)  We have
defined $\alpha = e^2/4\pi$.  
It is widely believed that the radiation reaction is given by
the Lorentz-Dirac formula\cite{Lorentz,Dirac}.
The corresponding four-force $K^\mu$ is given by
\begin{equation} 
K^{\mu} = \frac{2\alpha}{3} \left( - a^2 u^{\mu}
+ \frac{d^2 u^{\mu}}{d\tau^2}\right).  \label{AL}
\end{equation}
One can readily verify that $u^\mu K_\mu = 0$.  (This is necessary
for consistency of Newton's equation $m\dot{u}^\mu = K^\mu$, 
where $m$ is the mass of the charge.)
Moreover, if the acceleration
lasts only for a finite time, the total energy-momentum
lost due to this four-force agrees with the total energy-momentum radiated
in the form of
electromagnetic radiation according to the Larmor formula (\ref{Larmor}).

Although the Lorentz-Dirac formula (\ref{AL})
is widely accepted, it is known to have 
some unphysical features. (See, e.g., Refs.\ \cite{Jackson,IZ}.)  
Let us consider one-dimensional motion and let
$u^\mu = (u^0, u^1) = (\cosh \beta, \sinh \beta)$.
Then the Lorentz-Dirac force is
$K^{\mu} = \frac{2}{3}\alpha(\sinh \beta, \cosh \beta)\ddot{\beta}$. 
Newton's equation reads
\begin{equation}
\frac{d \beta}{d\tau} = \frac{2\alpha}{3m}\frac{d^2 \beta}{d\tau^2}
+F_{\rm ext}(\tau),
\label{2D}
\end{equation}
where $F_{\rm ext}(\tau)$ represents an external force.  
This equation allows runaway solutions.  For example, if 
$F_{\rm ext}(\tau) = 0$, then
$\beta (\tau) = \beta_0\exp( 3m\tau/2\alpha)$ is a solution.
Runaway solutions can be excluded by
rewriting Eq.\ (\ref{2D}) as follows:
$$
m\frac{d\beta}{d\tau} = \int_0^{\infty} e^{-s}
F_{\rm ext}(\tau +\tau_0 s)ds,
$$
where $\tau_0 = 2\alpha/3m$.  However, this equation
violates causality.  This causality violation is usually dismissed as
unimportant\cite{Jackson} because its time scale $\tau_0$ is very small.
(It is of order $10^{-24}$ sec for electrons.)  This nevertheless casts some
doubts on the validity of the Lorentz-Dirac theory.  In this Letter we will
show that in fact the Lorentz-Dirac theory does not describe the classical
limit of the quantum theory if the acceleration is due to a static potential
varying in one dimension, say in the $z$-direction.  
(Moniz and Sharp concluded that the Lorentz-Dirac formula is reproduced
in the classical limit by studying the Heisenberg equations\cite{Sharp}.
It is difficult to find the reason
why we disagree since our method is quite different
from theirs.)

Once we start considering a quantum system with a static potential,
a natural alternative classical theory emerges.  
The $z$-component
of the momentum is not conserved in the rest frame of the potential. 
However, the
energy is conserved.  Hence, it is natural to expect that the correct classical
limit is obtained by assuming that the energy is lost at each time according
to the Larmor formula (\ref{Larmor}), 
disregarding the momentum conservation in the
$z$-direction.  We call this theory
the Larmor theory in this Letter.
We show that the Larmor theory is likely to be the correct classical theory.
We use natural units $\hbar = c = 1$ unless otherwise stated throughout
this Letter. 

We consider a charged particle moving in the $z$-direction under
the influence of a potential $V(z)$ which is constant if $|z|$ is sufficiently
large. (Thus, the acceleration lasts only for a finite time.)  Let us first
clarify how the Lorentz-Dirac and Larmor theories
differ. 


We work in the 
non-relativistic approximation.
Let us assume that the kinetic energy is dominant
so that the particle always moves in the positive $z$-direction. 
The energy is defined as
$E = \frac{1}{2}m v^2 + V(z)$, where $v$ is the velocity of the particle.
The rate of energy change in the Lorentz-Dirac theory reads
\begin{equation}
\frac{dE_D}{dt} = \frac{2}{3}\alpha \ddot{v}v, \label{ABRA}
\end{equation}
where $\ddot{v} = d^2 v/dt^2$.  On the other hand, 
according to the Larmor theory this quantity is
\begin{equation}
\frac{dE_L}{dt} = -\frac{2}{3}\alpha \dot{v}^2. \label{SECOND}
\end{equation}
To first order in $\alpha$, one can let $v$ on the right-hand sides of
Eqs.\ (\ref{ABRA}) and (\ref{SECOND}) be the velocity without any
radiation reaction.  
By imposing the condition that $E_D = E_L$ for $t\to -\infty$
one has
\begin{equation}
E_D - E_L = \frac{\alpha}{3}\frac{d\ }{dt}v^2. \label{EAEL}
\end{equation}
Note here that this quantity vanishes for $t \to +\infty$ because
$V(z)$ is assumed to become constant for large $|z|$.  Thus, the two
theories give the same final velocity.

Now, let $z$ denote the position without the
radiation reaction. Let $\Delta_D v$ and $\Delta_D z$ be the corrections
to the velocity and position
at the lowest order in $\alpha$ according to the Lorentz-Dirac theory and
$\Delta_L v$ and $\Delta_L z$ be the corresponding quantities for the
Larmor theory.  Then, 
using $V'(z) = -m\dot{v}$, $\Delta_D v = (d/dt)(\Delta_D z)$ and
$\Delta_L v = (d/dt)(\Delta_L z)$, one finds from Eq.\ (\ref{EAEL}) 
$$
m\frac{d\ }{dt}\left(\frac{\Delta_D z - \Delta_L z}{v}\right)
= \frac{2\alpha}{3}\frac{d\ }{dt}\ln v.
$$
Let the initial and final velocities without the radiation reaction be
$v_i$ and $v_f$, respectively.  Then, for $t \to +\infty$, one has 
$$
\Delta_D z - \Delta_L z = \frac{2\alpha}{3m}v_f\ln\frac{v_f}{v_i}.
$$
Now, let $a(t)$ be the acceleration.
Then, to lowest order in $a(t)$ one finds
\begin{equation}
\Delta_D z - \Delta_L z
= \frac{2\alpha}{3m}(v_f - v_i) = \frac{2\alpha}{3m}\int_{-\infty}^{+\infty}
a(t)\, dt. \label{Difference}
\end{equation} 
Thus, the two theories differ in the position shift, and the difference is
of first order in acceleration.

Next let us derive the formula for the position shift at $t =0$ for the
Larmor theory, assuming
that $a(t) = 0$ for $t > 0$ for simplicity. 
The energy formula (\ref{SECOND}) gives
$$
\frac{d\ }{dt}(v\Delta_L v - \dot{v}\Delta_L z) = -\frac{2\alpha}{3m}
a(t)^2. 
$$
{}From this we find the position shift at $t=0$ to lowest order in $a(t)$ as
\begin{eqnarray}
(\Delta_L z)_0 & = & -\frac{2\alpha}{3\bar{p}}\int_{-\infty}^0
\left( \int_{-\infty}^{t}a(t')^2 dt'\right) dt \nonumber \\
 & = &
\frac{2\alpha}{3\bar{p}}
\int_{-\infty}^{+\infty} t\,a(t)^2 dt, \nonumber
\end{eqnarray}
where we have replaced $mv$ by the average momentum $\bar{p}$.
Define $\hat{a}(k)$ to
be the Fourier transform of $a(t)$, i.e.,
$$
\hat{a}(k) = \int_{-\infty}^{+\infty} dt\, a(t)e^{ikt}.
$$
Then we find
\begin{equation}
(\Delta_L z)_0 = -\frac{2i\alpha}{3\bar{p}}
\int_{-\infty}^{+\infty} \frac{dk}{2\pi}\,
\overline{\hat{a}(k)}\frac{d\ }{dk}\hat{a}(k).  \label{Fourier}
\end{equation}
Note that this position shift is of second order in acceleration. 
Since the difference in the position shift for $t \to +\infty$
given by Eq.\ (\ref{Difference}), which
equals that at $t=0$ if $a(t) = 0$ for $t> 0$, is of first order, 
the position shift $(\Delta_D z)_0$ at $t=0$
according to the Lorentz-Dirac theory is of first order in acceleration.
It will be shown
next that the position shift is of second order in acceleration in quantum
theory and, therefore, that the Lorentz-Dirac theory gives 
an incorrect classical limit.


The quantum system corresponding to the classical one considered above 
is given, to lowest non-trivial order in $e$, by the following Hamiltonian:
\begin{equation}
H = \frac{\hat{{\bf p}}^2}{2m} + V(z) + H_{\rm em} - \frac{e}{m}{\bf A}\cdot 
\hat{{\bf p}},
\label{Hamil}
\end{equation}
where $H_{\rm em}$ is the Hamiltonian describing the free electromagnetic
field and where $\hat{{\bf p}} = -i\partial/\partial {\bf x}$. 
The relevant Hilbert space is the tensor product of the space of
non-relativistic one-particle Schr\"odinger wave functions and the Fock
space of photons.  The underlying theory is the 
scalar QED with an external
potential for the charged scalar field. 
(See, e.g., Ref.\ \cite{Schiff} for the non-relativistic version of this
theory.)
One arrives at the Hamiltonian 
(\ref{Hamil}) by adopting 
the Coulomb gauge and dropping the terms of second order
in $e$.\footnote{The term proportional to 
$e^2A^{\mu}A_{\mu}\phi^\dagger\phi$ 
contributes only to the mass renormalization at
the lowest order. The term representing the Coulomb interaction produces
a contribution corresponding to Coulomb scattering of the charged particle
off a vacuum loop under the influence of $V(z)$.  This contribution vanishes
if one assumes that the potential $V(z)$ is the same (including the sign)
for both the positively and negatively charged particles.  We will make this
assumption here.}  

First let us consider wave functions evolved in time by the 
Hamiltonian
$H_0 = \hat{{\bf p}}^2/2m + V(z)$.
They represent accelerated particles without radiation reaction.
Let $V(z) = V_0 = {\rm const.}$ for sufficiently large and negative
values of $z$ and $V(z) = 0$ for sufficiently large and positive values of
$z$.  Writing the solution to the time-independent Schr\"odinger
equation as $\phi_P(z)\exp(iP_x x + iP_y y)$, where the energy is
$E = (P^2 + P_x^2 + P_y^2)/2m$, one finds
$$
\left[ -\frac{1}{2m}\frac{d^2\ }{dz^2} + V(z)\right]\phi_{P}(z) 
= \frac{P^2}{2m}\phi_{P}(z).
$$
Let us assume that $P^2/2m \gg |V(z)|$ and that $V(z)$ is smooth enough
so that the WKB approximation is applicable.  The right-moving solutions
$\phi_P(z)$ are given by
$$
\phi_{P}(z) = \kappa(z)^{-1/2}\exp\left[ 
i \int_0^{z} \kappa(z)\,dz\right],
$$
where $\kappa(z) = [P^2 - 2mV(z)]^{1/2}$. 

Let the initial state be
$\Psi_0 = \psi_0\otimes |0\rangle$,
where $|0\rangle$ is the vacuum state of the electromagnetic field and
$\psi_0$ is a wave packet of the charged particle given by 
$$
\psi_0 = \int\frac{dPd{\bf P}_{\perp}}{\sqrt{(2\pi)^3}}
f(P,{\bf P}_{\perp})
\sqrt{P}\,\phi_{P}(z)
e^{i{\bf P}_{\perp}\cdot {\bf x}_{\perp} - iE(P,{\bf P}_{\perp})t}
$$
Here, ${\bf P}_{\perp} = (P_x, P_y)$,
$E(P,{\bf P}_{\perp}) = (P^2 + {\bf P}_{\perp}^2)/2m$, 
${\bf x}_{\perp} = (x,y)$ and
$\int dPd^2{\bf P}_{\perp}|f(P,{\bf P}_{\perp})|^2 = 1$.
The function $f(P,{\bf P}_{\perp})$ is assumed to be non-zero only for
the values of $P$ much larger than $|2mV(z)|^{1/2}$.
The wave packet is assumed to be sharply peaked about a classical trajectory
near the region with $V'(z) \neq 0$.
The electromagnetic field ${\bf A}({\bf x})$ in the Schr\"odinger picture
can be expanded as
$$
{\bf A}({\bf x}) = \sum_{j = 1}^2
\int\frac{d^3{\bf k}}{\sqrt{(2\pi)^3 2k}}
\left[ \mbox{\boldmath $\epsilon$}^{(j)} a_{{\bf k}j}e^{i{\bf k}\cdot {\bf x}}
 + \mbox{\boldmath $\epsilon$}^{(j)} a_{{\bf k}j}^{\dagger}
e^{-i{\bf k}\cdot {\bf x}}\right],
$$
where \mbox{\boldmath $\epsilon$}$^{(j)}$ are the (real)
polarizations of the photon satisfying 
\mbox{\boldmath $\epsilon$}$^{(j)}\cdot {\bf k} = 0$ and
where $k = |{\bf k}|$. 
The annihilation and creation operators $a_{{\bf k}j}$ and 
$a_{{\bf k}j}^{\dagger}$ satisfy the usual commutation relations
$[a_{{\bf k'}j'},a_{{\bf k}j}^{\dagger}] 
= \delta_{j'j}\delta({\bf k}'-{\bf k})$ and
$[a_{{\bf k'}j'},a_{{\bf k}j}] =   
[a_{{\bf k'}j'}^{\dagger},a_{{\bf k}j}^{\dagger}] = 0$.

By straightforward application of the time-dependent perturbation theory
with the interaction Hamiltonian
$H_I = -(e/m){\bf A}\cdot \hat{{\bf p}}$, one finds the first-order component
in the final state as 
\begin{eqnarray}
\psi_1
& = & e \sum_{j=1}^2 \int d^2{\bf p}_{\perp}dp
\int \frac{d^3{\bf k}}{\sqrt{(2\pi)^3 2k}} 
\sqrt{\frac{p}{P}}f(P,{\bf P}_{\perp})|p{\bf p}_\perp{\bf k}j\rangle
\nonumber \\
 & & \times 
\int dz'\overline{\phi_{p}(z')}[ i\mbox{\boldmath $\epsilon$}^{(j)}\cdot
{\bf P}_{\perp}\,\phi_{P}(z') + \epsilon_z^{(j)}\,\phi'_{P}(z')]e^{ik_z z'}, 
\nonumber
\end{eqnarray}
where
$$
|p{\bf p}_{\perp}{\bf k}j\rangle
= 
\sqrt{p}\,\phi_{p}(z)e^{i{\bf p}_{\perp}\cdot{\bf x}_{\perp}
- iE(p,{\bf p}_{\perp})t} \otimes a_{{\bf k}j}^{\dagger}|0\rangle,
$$
${\bf P}_{\perp} = {\bf p}_{\perp} + {\bf k}_{\perp}$ and
$P = \sqrt{p^2 + {\bf p}_{\perp}^2
- {\bf P}_{\perp}^2 + 2mk}$.

At the lowest order in the WKB approximation, one has
$\phi_P'(z) \approx i[P^2 - 2mV(z)]^{1/2}\phi_P(z)$.
By choosing the wave packet such that $|{\bf P}_{\perp}| \ll P$ if
$f(P,{\bf P}_{\perp}) \neq 0$, the term proportional to 
$\mbox{\boldmath $\epsilon$}^{(j)}\cdot {\bf P}_{\perp}$ can be neglected.
One can then choose the two polarizations so that $\epsilon_z^{(2)} =0$.
Thus, one has
$$
\psi_1
 \approx  i e
\int \frac{d^2{\bf p}_{\perp}dp}{\sqrt{(2\pi)^3}}
\int \frac{d^3{\bf k}}{\sqrt{(2\pi)^3 2k}} 
\epsilon_z^{(1)}
I_{pk} f(P,{\bf P})|p{\bf p}_{\perp}{\bf k}j\rangle,
$$
where 
\begin{eqnarray}
I_{pk} & = & \sqrt{\frac{p}{P}}\int_{-\infty}^{+\infty}
dz\, \sqrt{P^2 - 2mV(z)}\,
\overline{\phi_p(z)}\phi_P(z)e^{-ik_z z} \nonumber \\
 & \approx &
\int_{-\infty}^{+\infty}
\exp \left[
i \int_0^z 
g(z') dz' -ik_z z\right].
\nonumber
\end{eqnarray}
We have defined
\begin{equation}
g(z) = 
[P^2 - 2mV(z)]^{1/2} - [p^2 - 2mV(z)]^{1/2}.
\label{gzed}
\end{equation}
It is assumed here that
the WKB approximation is applicable to the final-state wave functions as well.
This assumption is valid unless the photon carries away most of the kinetic
energy of the charged particle.  The phase space for such final states
can be made arbitrarily insignificant by increasing $P$. 
Hence, it is reasonable to
assume that the contribution from such final states can be neglected.

The integral $I_{pk}$ can be rewritten
by integrating by parts and dropping the surface terms as
$$
I_{pk} = -i\int_{-\infty}^{+\infty}dz\,\frac{g'(z)}{[g(z)-k_z]^2}
\exp\left[ i\int_0^z g(z')dz'\right]e^{- ik_z z}.
$$
(The surface terms can be dropped
because we are dealing with a wave packet here, and
the final result would be the same if we introduced a damping factor which
made the surface terms vanish.)
Now, the function $g(z)$
in Eq.\ (\ref{gzed}) is multiplied by $\hbar^{-1}$ if
one restores $\hbar$. Hence, in the classical limit $\hbar \to 0$, 
the integral $I_{pk}$ tends to zero due to rapid oscillations if
$P\neq p$.
Since we are interested
in the classical limit, we
make the approximation $P \approx p$.  Now, by using
conservation of transverse momentum,
${\bf P}_{\perp} = {\bf p}_{\perp} + {\bf k}_{\perp}$, one finds 
$|{\bf P}_{\perp}^2-{\bf p}_{\perp}^2|
\leq k(|{\bf P}_{\perp}|+|{\bf p}_{\perp}|)$.
Note that $m \gg |{\bf P}_{\perp}|+|{\bf p}_{\perp}|$ because the charged
particle is assumed to be non-relativistic. Therefore, we conclude that
$|{\bf P}_{\perp}^2 - {\bf p}_{\perp}^2| \ll 2mk$.
{}From $P = \sqrt{p^2 + {\bf p}_{\perp}^2 - {\bf P}_{\perp}^2 + 2mk}$ 
we have $P \approx [p^2 + 2mk]^{1/2}$.
Thus, in the non-relativistic
and classical regime,
one can make the following approximation:
$$
g(z) = [P^2 - 2mV(z)]^{1/2} - [p^2 - 2mV(z)]^{1/2}
\approx \frac{k}{v_p(z)},
$$
where $v_p(z) = [p^2 - 2mV(z)]^{1/2}/m$ 
is the velocity of the classical particle with final momentum
$p$ in the $z$-direction. By using this approximation 
one finds
$$
I_{pk} = i \int_{-\infty}^{+\infty}
dz \frac{kv'_p(z)e^{-ik_z z}}{[k-k_zv_p(z)]^2}
\exp \left[ i\int_0^{z}\frac{k}{v_p(z')}dz'\right].
$$
Noting that $k \gg k_z v_p(z)$ because the particle is non-relativistic,
one finds
$$
I_{pk} \approx \frac{i}{k}\int_{-\infty}^{+\infty}
dt\, a_p(t)e^{ikt} = \frac{i}{k}\hat{a}_p(k),
$$  
where $a_p(t)$ is the acceleration of the classical particle with 
the final momentum $p$, with $t = 0$ at $z=0$.  Thus, the first-order part
of the final-state wave function in the momentum representation is
$$
\hat{\psi}_1 \approx -e\epsilon_z^{(1)} 
\frac{\hat{a}_p(k)}{\sqrt{(2\pi)^3 2k^3}}
f(p,{\bf p}_{\perp}).
$$
We have let $P \approx p$ and ${\bf P}_{\perp} \approx {\bf p}_{\perp}$.
The probability of emission is obtained by integrating $|\hat{\psi}_1|^2$
over $p$, ${\bf p}_\perp$ and ${\bf k}$.
Here, we note that the amplitude of emission is proportional to the 
acceleration.  Hence, the position shift must be at least of second order
in acceleration because the probability is proportional to the amplitude
squared. Thus, the Lorentz-Dirac theory cannot give the correct classical
limit.  

By recalling that 
$\int dp d^2{\bf p}_{\perp}|f(p,{\bf p}_{\perp})|^2 = 1$,
the emission probability is found to be 
$$
{\cal P} = \frac{2e^2}{3}\int\frac{d^3{\bf k}}{(2\pi)^3 2k^3}
|\hat{a}_p(k)|^2 = \frac{4\alpha}{3}\int_0^{\infty} \frac{dk}{2\pi k}
|\hat{a}_p(k)|^2,
$$
where $p$ is taken to be the average value for the wave packet.  We have used
the fact that
the average of $(\epsilon^{(1)}_z)^2$ is 2/3.
The probability ${\cal P}$ is infrared divergent if
$\hat{a}_p(0) \neq 0$, i.e., 
if the initial and final velocities are different.
The expected energy of the emitted photon is
$$
E = \frac{e^2}{3\pi}\int_0^{\infty} \frac{dk}{2\pi}|\hat{a}_p(k)|^2
= \frac{2\alpha}{3} \int_{-\infty}^{+\infty} dt\,a(t)^2.
$$
Thus, the classical Larmor formula (\ref{Larmor}) is reproduced. 

Now, the position operator (in the momentum space)
of the final-state wave packet evolved back to
$t=0$ can be
approximated by $i\partial/\partial p$. 
(The time $t=0$ is automatically picked out because the phase factor 
describing the time dependence of the wave function is chosen to be one at
$t=0$.) 
Let us concentrate on the sector with ${\bf p}_{\perp}$ fixed and suppress
${\bf p}_{\perp}$. 
The contribution to the expectation value of $i\partial/\partial p$
from the tree diagram of first order
in $e$ is given by
\begin{eqnarray}
(z)_0 & = & \int
d^3{\bf k} \langle \hat{\psi}_1, \hat{\psi}_1\rangle \nonumber \\
 & = &  \frac{4e^2}{3}
\int\frac{d^3{\bf k}}{(2\pi)^3 2k^3} \langle
\hat{a}_p(k)f(p), \hat{a}_p(k)f(p)\rangle.  \nonumber
\end{eqnarray}
where
$$
\langle A, B\rangle = \frac{i}{2} \int dp \left[ 
\overline{A}\frac{\partial B}{\partial p}
- \frac{\partial \overline{A}}{\partial p}B\right].
$$
The acceleration is given by $- V'(z)/m$ as a function of $z$, 
and its $p$ and
$t$ dependence is only
through $z = pt/m$.  Thus, one has
$[t(\partial/\partial t) - p(\partial/\partial p)]a_p(t) = 0$.
Hence,
$$
\frac{\partial\ }{\partial p}\hat{a}_p(k)
= - \frac{1}{p}\hat{a}_p(k) 
- \frac{k}{p}\frac{\partial\ }{\partial k}\hat{a}_p(k)\,.
$$
By using this formula and the fact that
$\hat{a}_p(-k) = \overline{\hat{a}_p(k)}$ (because $a_p(t)$ is real), one
obtains
\begin{eqnarray}
(z)_0 & =  & \frac{4\alpha}{3}\langle f(p),f(p)\rangle 
\int_0^{\infty}\frac{dk}{2\pi k}
|\hat{a}_p(k)|^2 \nonumber \\
 & &  - \frac{2i\alpha}{3}\int_{-\infty}^{+\infty}\frac{dp}{p}
|f(p)|^2 \int_{-\infty}^{+\infty}
\frac{dk}{2\pi}
\overline{\hat{a}_p(k)}\frac{\partial\ }{\partial k}
\hat{a}_p(k). \nonumber 
\end{eqnarray}
The first term is the expectation value of the operator
$i\partial/\partial p$ (without radiation reaction)
times the probability of emission. 
This term will be
cancelled by the ``one-loop'' correction 
without emission at this order due to unitarity.
This means that
the expectation value of the final-state position {\it shift} evolved
back to $t=0$ is given by
$$
(\Delta z)_0 = - \frac{2i\alpha}{3}\int_{-\infty}^{+\infty}\frac{dp}{p}
|f(p)|^2 \int_{-\infty}^{+\infty}
\frac{dk}{2\pi}
\overline{\hat{a}_p(k)}\frac{\partial\ }{\partial k}
\hat{a}_p(k).  
$$
It is clear that one obtains Eq.\ (\ref{Fourier}) in the limit 
where the wave packet becomes concentrated at a point in the
momentum space. Thus, the Larmor theory gives the correct position shift.


In summary, it was shown in this Letter that the Lorentz-Dirac
radiation reaction formula does not give the correct classical limit of
the quantum mechanical system with a static potential, by demonstrating
that the position shift due to
that formula is incorrect.  
It was also shown that the correct classical limit of the position shift
is obtained by assuming
that the energy is lost according to the Larmor formula, disregarding
momentum conservation.


\acknowledgments

The author thanks Chris Fewster,
Bernard Kay and George Matsas for useful
discussions. 


\end{document}